\documentclass[a4paper,12pt]{article}
\usepackage{graphicx,epsf}

\usepackage{colordvi}
\begin{document}
\title{Inflation, quantum fluctuations and cosmological 
perturbations}
\author{David Langlois \\
GRECO, Institut d'Astrophysique de Paris (CNRS) \\
98bis Boulevard Arago, 75014 Paris, France}
\date{\today} 
\maketitle

\def\beq{\begin{equation}}
\def\eeq{\end{equation}}

\def\h{{\cal H}}

\def\d{\delta}
\def\B{{\bar B}}
\def\E{{\bar E}}
\def\V{{\cal V}}
\def\P{{\cal P}}
\def\R{{\cal R}}
\def\p{{\bf p}}
\def\D{D}

\def\L{{\rm L}}
\def\k{{\vec k}}
\def\x{{\vec x}}

\def\mP{m_P}  

\begin{abstract}
These lectures  are intended to 
give a  pedagogical introduction to the main current picture of the 
very early universe. After  elementary reviews of general relativity 
and of the standard Big Bang model, the following subjects 
are discussed:
inflation, the classical 
relativistic theory of cosmological perturbations and  
the generation of perturbations from scalar field quantum fluctuations 
during inflation.  
\end{abstract}

\section{Introduction}
The purpose of these lectures is to give an introduction to the 
present {\it standard picture of the early universe}, which 
{\it complements} the older standard Big Bang model. 
These notes are intended for non-experts on this subject.
 They start with a very short introduction to General Relativity, 
on which  modern cosmology is based, followed by an 
elementary review   of the standard Big Bang model. 
We then discuss the limitations of this model
and enter into the main subject  of these lectures: {\it inflation}.

Inflation was initially invented to solve some of the problems
of the standard Big Bang model and to get rid of unwanted relics  
generically predicted by high energy models. It turned out that 
inflation, as it was realized later, could also solve an additional 
puzzle of the standard model, that of the generation of the cosmological 
perturbations. This welcome surprise put inflation on a rather 
firm footing, about 
twenty years ago. Twenty years later, inflation is still alive, in 
a stronger position than ever because its few competitors have been 
 eliminated as 
 new cosmological observations have accumulated during the last few years.

\section{A few elements on general relativity and cosmology}
Modern cosmology is based on Einstein's theory of general relativity. It is 
thus useful, before discussing the early universe, to recall 
a few notions and useful formulas from this theory. Details can 
be found in  standard textbooks on general relativity (see e.g. \cite{GR}).
In the framework
of general relativity, the spacetime geometry is defined by 
a {\it metric}, a symmetric tensor with two indices, whose 
components in a coordinate system $\{x^\mu\}$ ($\mu=0,1,2,3$)  
will be denoted  $g_{\mu\nu}$.
The square of the ``distance'' between two neighbouring points of spacetime 
 is 
given by the expression 
\beq
ds^2=g_{\mu\nu}dx^\mu dx^\nu.
\eeq
We will use the signature  $(-,+,+,+)$. 

In a coordinate change $x^\mu\rightarrow \tilde x^\mu$, the new components
of the metric are obtained by using the standard tensor 
transformation formulas, namely
 \beq
\tilde g_{\mu\nu}={\partial x^\rho\over \partial \tilde x^\mu}
{\partial x^\sigma\over \partial \tilde x^\nu}g_{\rho\sigma}.
\eeq

One can define a {\it covariant derivative} associated to this metric, 
denoted $\D_\mu$, whose action on a tensor with, for example,
 one covariant index
and one contravariant index will be given by 
\beq
\D_\lambda T^{\mu}_{\hskip 2mm \nu}=\partial_\lambda T^{\mu}_{\hskip 2mm \nu}
+\Gamma^\mu_{\lambda\sigma}T^{\sigma}_{\hskip 2mm \nu}
- \Gamma^\sigma_{\lambda\nu}T^{\mu}_{\hskip 2mm \sigma}
\eeq
(a similar term must be added for each additional covariant or 
contravariant index),
where the  $\Gamma$ are the  {\it Christoffel symbols} (they are not 
tensors), defined by 
\beq
\Gamma^\lambda_{\mu\nu}={1\over 2}g^{\lambda\sigma}\left(\partial_\mu
g_{\sigma\nu}+\partial_{\nu}g_{\mu\sigma}-\partial_\sigma g_{\mu\nu}
\right).
\label{christoffel}
\eeq
We have used the notation $g^{\mu\nu}$ which corresponds, for the 
metric (and only for the metric), to the 
inverse of $g_{\mu\nu}$ in a  matricial sense, i.e. 
$g_{\mu\sigma}g^{\sigma\nu}=\delta_\mu^\nu$.

The ``curvature'' of spacetime is characterized by the {\it Riemann}
tensor, whose components can be expressed in terms of the 
Christoffel symbols according to the expression
\beq
R_{\lambda\mu\nu}^{\hskip 6mm \rho}=
\partial_\mu \Gamma_{\lambda\nu}^\rho
-\partial_\lambda \Gamma_{\mu\nu}^\rho+
\Gamma_{\lambda\nu}^\sigma\Gamma_{\sigma\mu}^\rho
- \Gamma_{\mu\nu}^\sigma\Gamma_{\sigma\lambda}^\rho.
\eeq

{\it Einstein's equations} relate the spacetime geometry to its matter content.
The geometry appears in Einstein's equations via the {\it Ricci tensor},
defined by 
\beq
R_{\mu\nu}=R_{\mu\sigma\nu}^{\hskip 6mm \sigma},
\eeq
and the  {\it scalar curvature}, which is the trace of the Ricci
tensor, i.e.
\beq
R=g^{\mu\nu}R_{\mu\nu}.
\eeq
The matter enters Einstein's equations via the {\it energy-momentum 
tensor}, denoted $T_{\mu\nu}$, whose time/time component 
corresponds to the energy density, the time/space components 
to the momentum density and the space/space component to the 
stress tensor. 
Einstein's equations then read
\beq
G_{\mu\nu}\equiv R_{\mu\nu}-{1\over 2}R\,  g_{\mu\nu}=8\pi G\,  T_{\mu\nu},
\label{einstein}
\eeq
where the tensor  $G_{\mu\nu}$ is called the  {\it Einstein tensor}.
Since, by construction, the  Einstein tensor satisfies the identity 
$\D_\mu G^{\mu}_{\, \nu}=0$,
any energy-momentum on the right-hand side of  Einstein's equation 
 must necessarily satisfy  the relation 
\beq
\D_\mu T^{\mu}_{\, \nu}=0, 
\label{DT0}
\eeq
which can be interpreted as a generalization, in the context of 
a curved spacetime, of the familiar conservation laws for energy and 
momentum.

The motion of a particule is described by its trajectory in spacetime, 
 $x^\mu(\lambda)$, where $\lambda$ is a parameter. A free particle, 
i.e. which does not feel any force (other than gravity), satisfies 
the {\it geodesic equation}, which reads
\beq
t^\sigma\D_\sigma t^\mu=0,
\eeq 
where $t^\mu=dx^\mu/d\lambda$ is the vector field tangent to the 
trajectory (note that the geodesic equation written in this form 
assumes that the parameter $\lambda$ is affine). Equivalently, the 
geodesic can be rewritten as 
\beq
{d^2x^\mu\over d\lambda^2}+\Gamma^{\mu}_{\rho\sigma}{dx^\rho\over d\lambda}
{dx^\sigma\over d\lambda}=0.
\label{geodesique}
\eeq
The geodesic equation applies both to
\begin{itemize}
\item  massive particles, in which case 
one usually takes as the parameter $\lambda$ the so-called {\it proper time}
so that the corresponding tangent vector 
$u^\mu$ is normalized:  $g_{\mu\nu}u^\mu u^\nu=-1$;

\item massless particles, in particular the photon, in which case 
the tangent vector, usually denoted $k^\mu$ is light-like, i.e. 
 $g_{\mu\nu}k^\mu k^\nu=0$.
\end{itemize}

Einstein's equations can also be obtained from a variational principle. The 
corresponding action reads
\beq
{\cal S}={1\over 16\pi G}\int d^4x\sqrt{-g}\left(R-2\Lambda\right)
+\int d^4x \sqrt{-g} {\cal L}_{mat}.
\eeq
One can check that the variation of this action with respect to the metric 
 $g_{\mu\nu}$ , upon using the  definition 
\beq
T^{\mu\nu}={2\over \sqrt{-g}}{\delta\left(\sqrt{-g}{\cal L}_{mat}\right)
\over{\delta g_{\mu\nu}}},
\label{def_Tmunu}
\eeq
indeed gives Einstein's equations 
\beq
G_{\mu\nu}+\Lambda g_{\mu\nu}=8\pi G\,  T_{\mu\nu}.
\eeq
This is a slight generalization of Einstein's equations (\ref{einstein}) that 
includes a {\it cosmological constant} $\Lambda$. 
It is worth noticing that the 
cosmological constant can also be interpreted as a particular 
energy-momentum tensor of the form
 $T_{\mu\nu}=-(8\pi G)^{-1}\Lambda g_{\mu\nu}$.

\subsection{Review of standard cosmology}

In this subsection,  the foundations of modern cosmology are briefly 
recalled. They 
follow from Einstein's equations introduced above and from a few 
hypotheses concerning spacetime and its matter content.
One of the essential assumptions of cosmology (so far 
confirmed by observations) 
is to consider, as a first approximation, the universe as 
being homogeneous and 
isotropic. Note that these symmetries define  implicitly a particular 
``slicing'' of spacetime, the corresponding space-like hypersurfaces 
being homogeneous and isotropic. A different slicing of the {\it same}
spacetime will give in general  space-like hypersurfaces that are not 
homogeneous and isotropic.

The above hypothesis turns out to be very restrictive and the only metrics
compatible with this requirement reduce to the so-called 
{\it  Robertson-Walker} metrics, which read in an appropriate coordinate 
system
\beq
ds^2=-dt^2+a^2(t)\left[{dr^2\over{1-\kappa r^2}}+
r^2\left(d\theta^2+\sin^2\theta d\phi^2\right)\right],
\label{RW}
\eeq
with $\kappa=0,-1,1$ depending on the curvature of spatial 
hypersurfaces: respectively flat, elliptic or hyperbolic.

The matter content compatible with the spacetime symmetries of homogeneity
and isotropy is necessarily described by an energy-momentum tensor of the 
form (in the same coordinate system as for the metric  (\ref{RW})):
\beq
T^{\mu}_{\, \nu}={\rm Diag}\left(-\rho(t), p(t),  p(t), p(t)\right).
\eeq
The quantity $\rho$ corresponds to an energy density and $P$ to a pressure.

One can show that the so-called comoving particles, i.e. those particles 
whose spatial coordinates are constant in time, satisfy the 
geodesic equation
 (\ref{geodesique}). 

\subsection{Friedmann-Lema\^\i tre equations}

Substituting the  Robertson-Walker metric (\ref{RW}) 
in Einstein's equations (\ref{einstein}), one gets the so-called 
 {\it  Friedmann-Lema\^\i tre equations}:
\begin{eqnarray}
\left({\dot a\over a}\right)^2 &=& {8\pi G\rho\over 3}- {\kappa\over a^2},
\label{friedmann}
\\
{\ddot a\over a} &=& -{4\pi G\over 3}\left(\rho+3 P\right).
\label{friedmann2}
\end{eqnarray}
An immediate consequence of these two equations is the {\it continuity 
equation}
\beq
\dot \rho+3H\left(\rho+p\right)=0,
\label{conservation}
\eeq
where $H\equiv \dot a/a$ is the {\it Hubble parameter}.
The continuity equation 
 can be also obtained directly from the energy-momentum conservation 
 $\D_\mu
T^{\mu}_{\, \nu}=0$, as mentioned before.

In order to determine the cosmological evolution, it is easier to combine
(\ref{friedmann}) with  (\ref{conservation}). 
Let us assume an equation of state for the cosmological matter of the form
$p=w\rho$ with
$w$ constant. This includes the two main types of matter that play an 
important r\^ole in cosmology:
\begin{itemize}
\item gas of relativistic particles, $w=1/3$;

\item non relativistic matter, $w\simeq 0$.
\end{itemize}
In these cases, the conservation equation  (\ref{conservation}) 
can be integrated to give
\beq
\rho\propto a^{-3(1+w)}.
\eeq
Substituting in (\ref{friedmann}), one finds,  for  $\kappa=0$,  
\beq
3{\dot a^2\over a^2}=8\pi G\rho_0\left(a\over a_0\right)^{-3(1+w)}, 
\eeq
where, by convention, the subscript '0' stands for {\it present}
quantities.
One thus finds $\dot a^2\propto a^{2-3(1+w)}$, which gives for the 
evolution of the scale factor
\begin{itemize}
\item in a universe dominated by  non relativistic matter
\beq
a(t)\propto t^{2/3},
\eeq
\item 
and in a universe dominated by  radiation
\beq
a(t)\propto t^{1/2}.
\eeq
\end{itemize}

One can also mention
 the case of a {\it cosmological constant}, which corresponds
to an equation of state $w=-1$ and thus implies 
an exponential evolution for the scale
factor
\beq
a(t)\propto \exp(Ht).
\eeq

More generally, when  several types of matter coexist with respectively 
 $p_{(i)}=w_{(i)}\rho_{(i)}$, it is convenient to introduce the 
dimensionless parameters 
\beq
\Omega_{(i)}={8\pi G \rho_0^{(i)}\over 3 H_0^2},
\eeq
which express the {\it present} ratio of  the energy density of some
given species with respect  to the so-called {\it critical energy density} 
$\rho_{crit}= 3 H_0^2/(8\pi G)$, which corresponds to the total energy
density for a flat universe.

One can then rewrite the first 
 Friedmann equation  (\ref{friedmann}) as 
\beq
\left({H\over H_0}\right)^2=\sum_i\Omega_{(i)}
\left(a\over a_0\right)^{-3(1+w_{(i)})}+\Omega_\kappa
\left(a\over a_0\right)^{-2},
\eeq
with $\Omega_\kappa=-\kappa/a_0^2H_0^2$, which implies that the 
cosmological parameters must satisfy the consistency relation
\beq
\sum_i \Omega_{(i)}+\Omega_\kappa=1.
\eeq
As for the second Friedmann equation (\ref{friedmann2}), it implies
\beq
{\ddot a_0\over a_0 H_0^2}=-{1\over 2}\sum_i \Omega_{(i)}(1+w_{(i)}).
\eeq
Present cosmological observations yield for the various parameters
\begin{itemize}
\item Baryons: $\Omega_b\simeq 0.05$,

\item Dark matter: $\Omega_d\simeq 0.25$,

\item Dark energy (compatible with a cosmological constant): 
$\Omega_{\Lambda}\simeq 0.7$,

\item Photons: $\Omega_\gamma\simeq 5\times 10^{-5} $.
\end{itemize}
Observations have not detected so far any deviation from flatness. 
Radiation is very subdominant today but extrapolating 
backwards in time, radiation was dominant in the past 
  since its energy density scales 
as $\rho_\gamma\propto a^{-4}$ in contrast with non relativistic matter 
($\rho_m\propto a^{-3}$). 
Moreover, since the present matter content seems dominated by dark energy 
similar to a cosmological constant ($w_\Lambda=-1$), this indicates that 
our universe is presently accelerating.

\subsection{The cosmological redshift}
An important consequence of the  expansion of the universe  is the 
{\it cosmological redshift} of photons. This is in fact how the expansion 
of the universe was discovered initially. 

Let us consider two light signals emitted by a comoving object 
at two successive instants
$t_e$ and $t_e+\delta t_e$, then received later at  
respectively $t_o$ and  $t_o +\delta t_o$ by a (comoving) observer.
One can always set the observer at the center of the coordinate system.
All light trajectories reaching the observer are then radial and one 
can write, using (\ref{RW})
\beq
\int_0^{r_e}{dr\over \sqrt{1-\kappa r^2}}=\int_{t_e}^{t_o}{dt\over a(t)}.
\eeq
The left-hand side being identical for the two successive trajectories,
the right-hand side must vanish, which yields
\beq
{\delta t_o\over a_o}-{\delta t_e\over a_e}=0.
\eeq
This implies for the frequencies measured at emission and 
at reception a  redshift  given by 
\beq
1+z\equiv {\nu_e\over \nu_o}={a_o\over a_e}.
\eeq

\subsection{Thermal history of the universe}
To go beyond a simply geometrical description of cosmology,
it is very fruitful to apply thermodynamics to  the matter content of the 
universe. One can then  define a temperature $T$ for the cosmological 
photons, not only when they are strongly interacting with ordinary matter
but also after they have decoupled because, 
with the  expansion, the thermal distribution for the gas of 
photons is unchanged  except for a global rescaling of the 
temperature so that $T$ essentially evolves as 
\beq
T(t)\propto {1\over a(t)}.
\eeq
This means that, going  backwards in time, the universe was hotter 
and hotter. This is the essence  of the hot Big Bang scenario.

 As the universe evolves,  the reaction rates between the various 
species  are modified. A detailed analysis of these changes enables 
to reconstruct the past thermal history 
of the universe. 
Two events in  particular play an essential  r\^ole because of 
their observational consequences:
\begin{itemize}

\item Primordial nucleosynthesis

Nucleosynthesis occured at a temperature 
around $0.1$ MeV, when the average kinetic energy  
became  sufficiently low so that nuclear binding was possible. Protons 
and neutrons could then combine, which lead to the production of 
light elements, such that Helium, Deuterium, Lithium, etc... Within 
the observational uncertainties, this 
scenario is  remarkably confirmed  by the present measurements. 

\item Decoupling of baryons and photons (or last scattering)

A more recent event is the so-called ``recombination'' of nuclei and 
electrons to form atoms. This occured  at a temperature of the order of 
the eV. Free electrons thus almost disappeared, which entailed an effective
decoupling of the cosmological photons and ordinary matter. What we 
see today as the Cosmic Microwave Background (CMB) is made of the fossil 
photons, which interacted  for the last time with matter at the last 
scattering epoch. The CMB represents a remarkable  observational 
tool for analysing the perturbations of the  early universe, as well as
for measuring the cosmological parameters introduced above.

\end{itemize}

\subsection{Puzzles of the standard Big Bang model}
The standard Big Bang model has encountered remarkable successes, in 
particular with the nucleosynthesis scenario and the prediction of the CMB, 
and it remains today a cornerstone in our understanding of the present 
and past universe.
However, a few intriguing facts  remain unexplained in the strict 
scenario of the standard Big Bang model and seem to necessitate a larger 
framework. We review below the main problems:

\begin{itemize}

\item Homogeneity problem

A first question is why the approximation of homogeneity and isotropy 
turns out to be so good. Indeed, inhomogeneities are 
unstable, because of gravitation, and they tend to grow with time. 
It can be verified for instance with the CMB that inhomogeneities were 
much smaller at the last scattering epoch than today. One thus expects that 
these homogeneities were still smaller further back in time.
How to explain a universe so smooth in its past ?

\item Flatness problem

Another puzzle lies in the (spatial) flatness of our universe. Indeed, 
Friedmann's equation (\ref{friedmann}) implies
\beq
\Omega-1\equiv {8\pi G \rho\over 3 H^2}-1={\kappa\over a^2 H^2}.
\eeq
In standard cosmology, the scale factor behaves like
$a\sim t^p$ with $p<1$  ($p=1/2$ for radiation and $p=2/3$ for 
non-relativistic  matter).
 As a consequence,  $(aH)^{-2}$  grows with time and 
  $|\Omega-1|$ must thus  diverge with time. Therefore, in the 
context of the standard model, the quasi-flatness observed today 
requires an extreme fine-tuning of $\Omega$ near $1$ in the early 
 universe.

\item Horizon problem

\begin{figure}
\begin{center}
\includegraphics[width=4.8in]{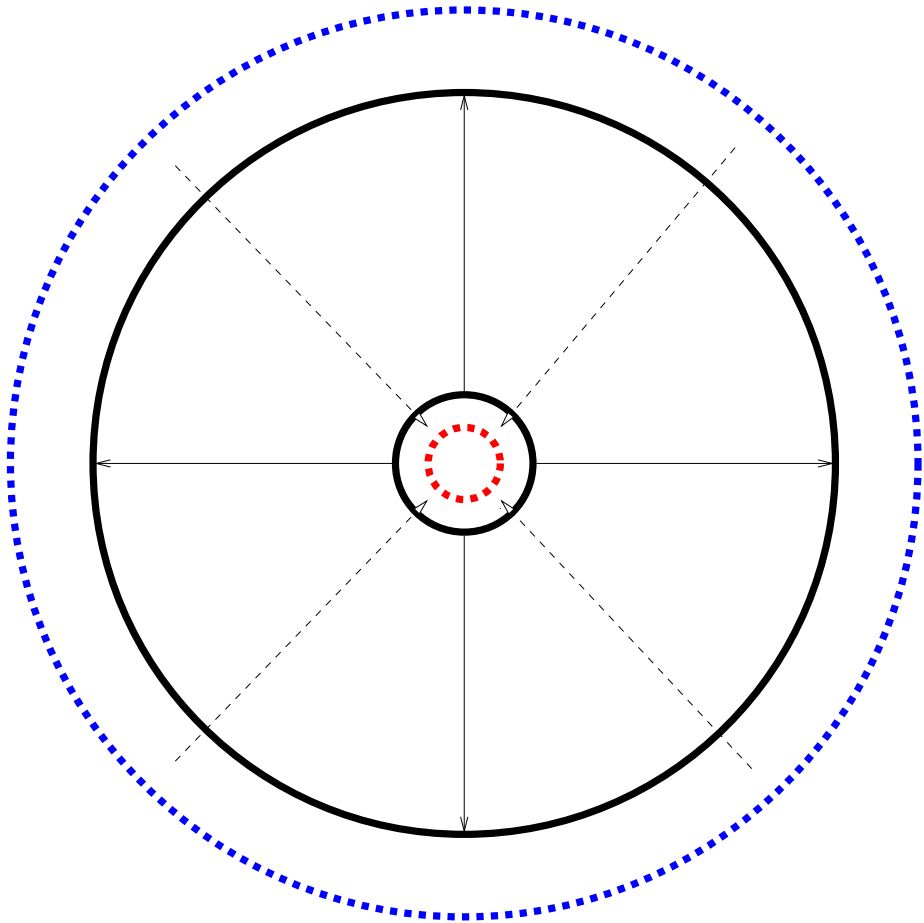}
\end{center}
\caption{Evolution of the comoving Hubble radius $\lambda_H=(aH)^{-1}$: during standard
cosmology, $\lambda_H$ increases (continuous lines), whereas during inflation
$\lambda_H$ shrinks (dashed lines).}
\end{figure}

One of the most fundamental problems in standard cosmology is certainly
the {\it horizon problem}.
The (particle) {\it horizon} is the maximal distance that can be covered 
by a light ray. 
For a light-like radial trajectory 
 $dr=a(t) dt$ and   the horizon is thus given by 
\beq
d_{H}(t)= a(t)\int_{t_i}^t{dt'\over a(t')}=a(t){t^{1-q}-t_i^{1-q}\over 1-q}, 
\eeq
where the last equality is obtained by assuming $a(t)\sim t^q$ and $t_i$ is
some initial time.

In standard cosmology ($q<1$), the integral converges in the limit 
 $t_i=0$ and the horizon has a finite size, of the order of the 
so-called Hubble radius $H^{-1}$:
\beq
d_H(t)={q\over 1-q}H^{-1}.
\eeq 
 
It also useful to consider 
 the {\it comoving Hubble radius}, $(aH)^{-1}$, which represents
the fraction of comoving space in causal contact.
One finds that it  {\it grows} with time, which means that the 
{\it fraction of the universe in causal contact increases with time} in 
the context of standard cosmology.
But the CMB tells us that the Universe was quasi-homogeneous 
at the time of last scattering on a scale encompassing 
 many regions a priori causally independent.
How to explain this ? 
\end{itemize}

A solution to the horizon problem and to the other puzzles 
is provided by the inflationary scenario, which 
we will examine in the next section. The basic idea
 is to invert the behaviour of the 
comoving Hubble radius, that is to make him {\it decrease} sufficiently in the very 
early universe. 
The corresponding condition is that 
\beq
\ddot a>0,
\eeq
i.e. that the Universe must undergo a {\it phase of acceleration}. 

\section{Inflation}
The broadest definition of inflation is that it corresponds to a phase
of acceleration of the universe,
\beq
\ddot a>0.
\eeq
In this broad sense, the current cosmological observations, if correctly
interpreted, mean that our present universe is undergoing 
 an inflationary phase. We are however interested here in an inflationary 
phase taking place in the very early universe, with different energy 
scales.

The Friedmann equations   (\ref{friedmann}) tell us that one can get 
acceleration  only if  the equation of state satisfies the condition 
\beq
P<-{1\over 3}\rho,
\eeq
condition which looks at first view   rather exotic.

A very simple example giving such an equation of state is 
a cosmological constant, corresponding to a cosmological fluid
with the equation of state
\beq
P=-\rho.
\eeq
However, a strict cosmological constant leads to 
exponential inflation {\it forever} which cannot be followed by 
a radiation or matter era.
Another possibility is a scalar field, which we discuss now in some 
details. 

\subsection{Cosmological scalar fields}

The dynamics  of a scalar field coupled to gravity is governed by the 
action 
\beq
\label{action_scalar_field}
S_\phi=\int d^4x\sqrt{-g}\left(-{1\over 2}\partial^\mu\phi\partial_\mu\phi
-V(\phi)\right).
\eeq
The corresponding energy-momentum tensor, which can be derived using  
(\ref{def_Tmunu}), is given by  
\beq
T_{\mu\nu}=\partial_\mu\phi\partial_\nu\phi-g_{\mu\nu}
\left({1\over 2}\partial^\sigma\phi\partial_\sigma\phi
+V(\phi)\right).
\label{Tscalarfield}
\eeq
If one assumes the geometry, and thus the matter,
 to be homogeneous and isotropic, 
then the energy-momentum tensor reduces to the perfect fluid form
with the energy density
\beq
\rho=-T_0^0={1\over 2}\dot\phi^2+V(\phi),
\eeq
where one recognizes the sum of a kinetic energy and 
of a potential energy, and the pressure
\beq
p={1\over 2}\dot\phi^2-V(\phi).
\eeq
The equation of motion for the scalar field is the Klein-Gordon 
equation, which is obtained by taking the variation of the above 
action (\ref{action_scalar_field}) 
with respect to the scalar field and which reads
\beq
\D^\mu\D_\mu\phi=V',
\eeq
in general and 
\beq
\ddot\phi+3H\dot \phi+V'=0
\eeq
in the particular case of a 
FLRW (Friedmann-Lema\^\i tre-Robertson-Walker) universe.

The system of equations governing the dynamics of the scalar field 
and of the geometry in a FLRW universe is thus given by
\begin{eqnarray}
& &H^2={8\pi G\over 3}\left({1\over 2}\dot\phi^2+V(\phi)\right), 
\label{e1}\\
& &\ddot\phi+3H\dot \phi+V'=0, 
\label{e2}\\
& & \dot H=-4\pi G\dot\phi^2.
\label{e3}
\end{eqnarray}
The last equation can be derived from the first two and is therefore
redundant. 

\subsection{The slow-roll regime}
The dynamical system (\ref{e1}-\ref{e3}) does not always give an accelerated
expansion but it does so in the so-called {\it slow-roll regime} when 
the potential energy of the scalar field dominates over its kinetic 
energy. 

More specifically, 
 the so-called 
{\it slow roll} approximation consists in neglecting 
the kinetic energy of the scalar field , $\dot \phi^2/2$ in (\ref{e1}) 
and the acceleration $\ddot\phi$ in the  Klein-Gordon equation 
(\ref{e2}). One then gets the simplified system
\begin{eqnarray}
& &H^2\simeq{8\pi G\over 3} V, 
\label{sr1}\\
& &3H\dot \phi+V'\simeq 0. 
\label{sr2}
\end{eqnarray}
Let us now examine in which regime this  approximation is valid. 
From
(\ref{sr2}), 
the velocity of the scalar field is given by
\beq
\dot\phi\simeq -{V'\over 3H}.
\label{phisr}
\eeq
Substituting this relation in the  condition $\dot \phi^2/2 \ll V$ yields 
the requirement:
\beq
\epsilon_V\equiv {\mP^2\over 2}\left({V'\over V}\right)^2 \ll 1,
\label{epsilon}
\eeq
where we have introduced the {\it reduced Planck mass}
\beq
m_P\equiv {1\over \sqrt{8\pi G}}.
\eeq
Similarly, the time derivative of (\ref{phisr}), using 
the time derivative of (\ref{sr1}), gives, 
after substitution in $\ddot \phi\ll V'$, the condition 
\beq
\eta_V\equiv \mP^2 {V''\over V}\ll 1.
\label{eta}
\eeq
In summary, the slow-roll approximation is valid when the 
two
 conditions $\epsilon_V, \eta_V \ll 1$ are satisfied, which means 
that the slope and the curvature of the potential, in Planck units, 
must be sufficiently small.

\subsection{Number of e-folds}
When working with a specific inflationary model, it is important to 
be able to relate the cosmological scales observed at the present time with 
the scales during inflation. For this purpose, one usally introduces 
the {\it number of e-foldings before the end of inflation}, denoted $N$,
and simply defined by 
\beq
N=\ln {a_{end}\over a},
\eeq
where $a_{end}$ is the value of the scale factor at the end of inflation
and $a$ is a fiducial value for the scale factor during inflation.
By definition, $N$ {\it decreases} during the inflationary phase 
and reaches zero at its end. 
In the slow-roll approximation, it is possible to express 
$N$ as a function of the scalar field. Since 
 $dN=-d\ln a=-H dt=-(H/\dot\phi) d\phi$, one easily finds, using
(\ref{phisr}) and (\ref{sr1}), that 
\beq
N(\phi)\simeq \int_\phi^{\phi_{end}}{V\over m_P^2 V'}d\phi.
\eeq
Given an explicit potential $V(\phi)$, one can in principle integrate 
the above expression to obtain $N$ in terms of $\phi$. This will be 
illustrated below for  some specific models.

Let us now discuss  the link  between $N$ and the present 
cosmological scales. Let us consider a given scale characterized by 
its comoving wavenumber $k=2\pi/\lambda$.
This scale crossed outside  the Hubble radius, 
during inflation, at an  instant $t_*(k)$ defined by 
\beq
k=a(t_*) H(t_*).
\eeq
To get a rough estimate  of the number of e-foldings of 
inflation that are needed 
to solve the horizon problem, let us first ignore the transition 
from a radiation era to a matter era and assume for simplicity that 
the inflationary phase was followed instantaneously 
by a radiation phase that has lasted   until now.
During the radiation phase, the comoving Hubble radius  $(aH)^{-1}$ 
increases like $a$. In order to solve the horizon problem, the increase
of the comoving Hubble radius during the standard evolution 
must be compensated by {\it at least}
a decrease of the same amount during inflation. 
Since the comoving Hubble radius 
roughly scales like $a^{-1}$ during inflation, the minimum amount 
of inflation is simply 
given by the number of e-folds between the end of inflation 
and today  
 $\ln(a_0/a_{end}) = 
\ln(T_{end}/T_0)\sim \ln(10^{29}(T_{end}/ 10^{16} {\rm GeV}))$, i.e. 
around 60 e-folds for a temperature $T\sim 10^{16} {\rm Gev}$ 
at the beginning of the radiation era. As we will see later, this 
energy scale is typical of inflation  in the simplest models.

This determines roughly the number of e-folds $N(k_0)$ 
between the moment when the scale corresponding to our present 
Hubble radius $k_0=a_0H_0$ exited the Hubble radius during inflation and 
the end of inflation.
 The other lengthscales  of cosmological interest are {\it smaller} than 
$k_0^{-1}$ and therefore exited the Hubble radius during inflation 
{\it after}  the scale $k_0$, whereas they entered the Hubble 
radius during the standard cosmological phase (either in the radiation era
for the smaller scales or in the matter era for the larger scales) 
{\it before} the scale $k_0$.

A more detailed calculation, which distinguishes between the energy
scales at the end of inflation and after the reheating, gives
for the number of e-folds between the exit of the mode $k$ and the end
of inflation 
\beq
N(k)\simeq 62 -\ln{k\over a_0 H_0}+\ln{V_k^{1/4}\over 10^{16}{\rm GeV}}
+ \ln{V_k^{1/4}\over V_{\rm end}^{1/4}}
+{1\over 3}\ln{\rho_{\rm reh}^{1/4}\over 
 V_{\rm end}^{1/4}}.
\eeq
Since the
 smallest  scale of cosmological relevance 
is  of the order of  $1$ Mpc, the range of cosmological 
scales  covers about $9$ e-folds.

The above number of e-folds is altered if one changes the thermal 
history of the universe between inflation and the present time 
 by including for instance
a period of so-called thermal inflation.

\subsection{Power-law potentials}
It is now time to illustrate all the points  discussed above 
with some specific potential. We consider first the case of power-law monomial 
potentials, of the form 
\beq
\label{pot_power-law}
V(\phi)={\lambda\over p}\mP^4 \left({\phi\over \mP}\right)^p,
\eeq
which have been abundantly used in the literature. In particular, 
the above potential includes the case of a free massive scalar field, 
$V(\phi)=m^2\phi/2$.
The slow-roll  conditions  $\epsilon\ll 1$ and  $\eta \ll 1$ 
both imply   
\beq 
\phi \gg p \  \mP,
\label{sr_pl}
\eeq
which means that the scalar field amplitude must be above the Planck mass
during inflation. 

After subsituting the potential (\ref{pot_power-law}) into the 
slow-roll equations of motion (\ref{sr1}-\ref{sr2}), 
one can integrate them explicitly to get 
\beq
\phi^{2-{p\over 2}}-  \phi_i^{2-{p\over 2}}
 =-{2\over 4-p}\, \sqrt{p\lambda\over 3}\, 
\mP^{3-{p\over 2}}\left(t-t_i\right)
\eeq
for $p\neq 4$ and
\beq
\phi=\phi_i\exp\left[-\sqrt{4\lambda\over 3}\,  \mP (t-t_i)\right]
\eeq
for $p=4$.

One can also express the scale factor as a function of the scalar field
(and thus as a function of time by substituting the above expression for
 $\phi(t)$) by using  $d\ln a/ d\phi=H/\dot \phi
\simeq-\phi/(p m_P^2)$. One finds 
\beq
a=a_{end}\exp\left[-{\left(\phi^2-\phi_{end}^2\right)\over 2p\,  m_P^2}\right].
\eeq
Defining the end of inflation by $\epsilon_V=1$, one gets 
$\phi_{end}=p \, \mP/\sqrt{2}$ and the number of e-folds is thus given by
\beq
\label{N_phi}
N(\phi)\simeq {\phi^2\over 2p \mP^2}-{p\over 4}.
\eeq
This can be inverted, so that
\beq
\phi(N)\simeq \sqrt{2Np}\, \mP,
\eeq
where we have ignored the second term of the right hand side of 
(\ref{N_phi}), in agreement with the condition (\ref{sr_pl}).

\subsection{Exponential potential}
If one considers a potential of the form
\beq
V=V_0 \exp \left(-\sqrt{2\over q}\, {\phi\over m_P}\right),
\eeq
then it is possible to find an {\it exact} solution (i.e. valid beyond 
the slow-roll approximation) of the system (\ref{e1}-\ref{e3}), with a
power-law   scale factor, i.e.
\beq
a(t)\propto t^q.
\eeq
The evolution of the scalar field is given by the expression
\beq 
\phi(t)={\sqrt{2q}\, m_P}\ln\left[\sqrt{V_0\over q(3q-1)} {t\over m_P}\right].
\eeq
Note that  one recovers the slow-roll approximation in the limit 
 $q\gg 1$, since the slow-roll parameters are given by  $\epsilon_V=1/q$ and 
$\eta_V=2/q$.

\subsection{Brief history of the inflationary models}
Let us now try to summarize in a few lines the history 
of inflationary models. The first model of inflation is usually traced back
to Alan Guth \cite{Guth:1980zm}
in 
1981, although  one can see a precursor in  the model  of 
Alexei Starobinsky \cite{Starobinsky:te}.
 Guth's model, which is named today {\it old inflation}
is based on a first-order phase transition, from a false vacuum with 
non zero energy, which generates  an exponential inflationary phase, into 
a true vacuum with zero energy density. The true vacuum phase appears 
in the shape of bubbles via quantum tunneling. 
The problem with this inflationary model is that, in order to get sufficient
inflation to solve the problems of the standard model mentioned earlier, 
the nucleation rate must be sufficiently small; but, then, the bubbles never 
coalesce because the space that separates the bubbles undergoes 
inflation and expands too rapidly.  Therefore, the first model of inflation 
is not phenomenologically viable.

After this first and unsuccessful attempt, a new generation of inflationary 
models appeared, usually denoted {\it new inflation} models 
\cite{new-inflation}. They rely 
 on a second order phase transition, based on thermal corrections of the 
effective potential and thus assume that the scalar field is in 
thermal equilibrium.

This hypothesis of thermal equilibrium 
was given up in the third generation of models, initiated 
by Andrei Linde, and whose generic name is 
{\it chaotic inflation} \cite{Linde:gd}.
This allows to use extremely simple potentials, quadratic or quartic, which 
lead to inflationary phases when the scalar field is displaced from the 
origin with values of the order of several Planck masses. This is sometimes
considered to be  
problematic from a particle physics point of view, as discussed briefly 
later.

During the last few years, there has been a revival of the inflation 
model building based on high energy theories, in particular in the 
context of supersymmetry. In these models, the value of the scalar field 
is much smaller than the Planck mass.

\subsection{The inflationary ``zoology''}

There exist many models of inflation. As far as single-field models 
are concerned (or at least {\it effectively} single field during inflation,
the hybrid models requiring a second field to end inflation as 
discussed below), it  is convenient to regroup them
into three broad categories:
\begin{itemize}
\item Large field models ($0<\eta\leq \epsilon$)

The scalar field is displaced from its stable minimum by $\Delta \phi\sim m_P$.
This includes the so-called chaotic type models with monomial potentials
\beq
V(\phi)=\Lambda^4\left({\phi\over \mu}\right)^p,
\eeq
or the exponential potential
\beq
V(\phi)=\Lambda^4 \exp\left(\phi/ \mu\right),
\eeq
which have already been discussed.

This category of models is widely used  in the literature because 
of the computational simplicity. But they are not always 
considered to be models which can be well motivated by   particle 
physics. The reason is the following. The generic potential 
for a scalar field will contain an infinite number of terms,
\beq
V(\phi)=V_0+{1\over 2}m^2\phi^2+{\lambda_3\over 3}\phi^3
+{\lambda_4\over 4}\phi^4+\sum_{d=5}^\infty m_P^{4-d}\phi^d,
\eeq
where the non-renormalizable  ($d>4$) couplings $\lambda_d$ are a priori 
of order $1$. When the scalar field is of order of a
few Planck masses,  one has no control on the form of the 
potential, and  all the non-normalizable terms must be taken into 
account in principle. 

In order to work with more specific forms for the 
potential,  inflationary model builders tend to 
concentrate on models where the scalar field amplitude is small with 
respect to the Planck mass, as those discussed just below.

\item Small field models ($\eta<0<\epsilon $)

In this type of models, 
the scalar field is rolling away from an unstable maximum of the potential.
This is a characteristic feature  of spontaneous symmetry breaking.
A typical potential is 
\beq
V(\phi)=\Lambda^4 \left[1-\left({\phi\over\mu}\right)^p\right],
\eeq
which can be interpreted as the lowest-order term in a Taylor expansion 
about the origin. Historically, this potential shape appeared in 
the so-called `new inflation' scenario.

A particular feature of these  models is that tensor modes 
are much more  suppressed with respect to scalar modes than in the 
large-field
models, as  it will be shown later.

\begin{figure}
\begin{center}
\includegraphics[width=4.8in]{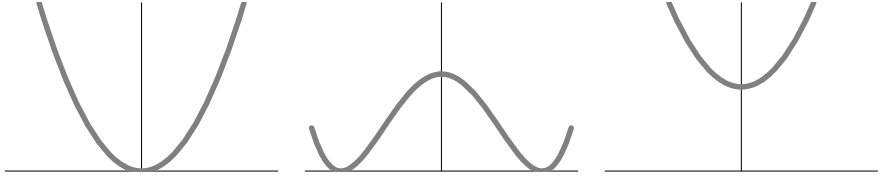}
\end{center}
\caption{Schematic potential for t
he three main categories of inflationary models: a. chaotic models 
b. symmetry breaking models; c. hybrid models}
\end{figure}

\item Hybrid models ($0<\epsilon<\eta$)

\begin{figure}
\begin{center}
\includegraphics[width=4.8in]{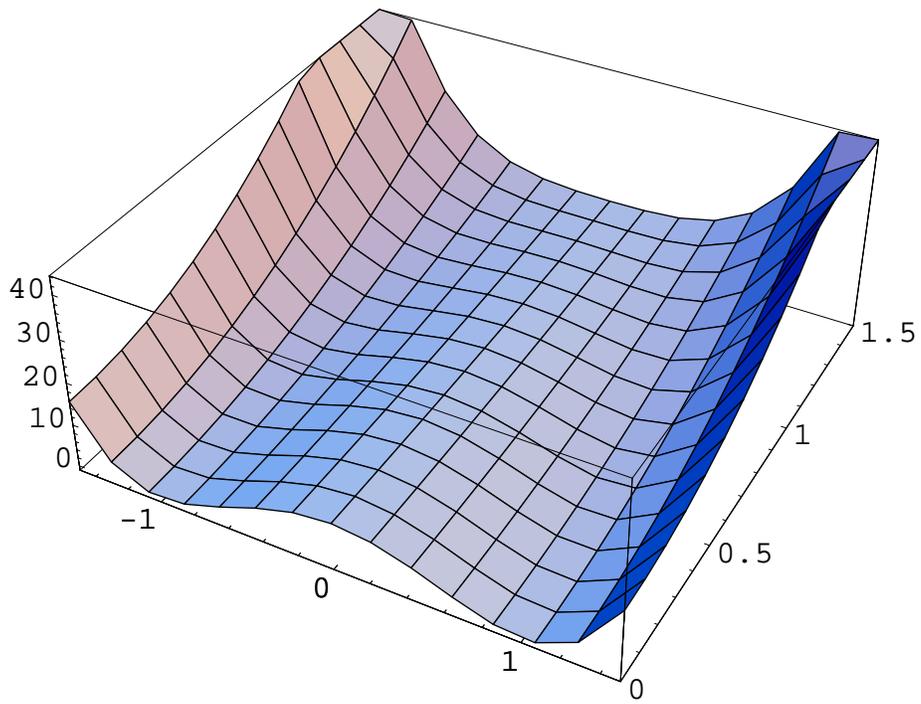}
\end{center}
\caption{Typical potential $V(\phi, \psi)$ for hybrid inflation.}
\end{figure}

This category of models,  which appeared rather recently, 
relies on the presence 
of two scalar fields: one plays the traditional r\^ole of the inflaton, 
while the other is  necessary to end inflation.

As an illustration, let us consider the original model of hybrid 
inflation \cite{hybrid} based on the potential 
\beq
V(\phi, \psi)={1\over 2}m^2\phi^2+{1\over 2}\lambda' \psi^2\phi^2+
{1\over 4}\lambda\left(M^2-\psi^2\right)^2.
\eeq
For values of the field $\phi$ larger than the critical value 
 $\phi_c=\lambda M^2/\lambda'$, the potential for $\psi$ has its minimum 
at $\psi=0$. This is the case during inflation. $\psi$ is thus
trapped in this minimum $\psi=0$, so that the effective potential 
for the scalar field $\phi$, which plays the r\^ole of the inflaton, is 
given by 
\beq
V_{eff}(\phi)=V_0+{1\over 2}m^2\phi^2,
\eeq
with $V_0=\lambda M^4/4$.
 During the inflationary phase, the field $\phi$ slow-rolls until it reaches 
the critical value  $\phi_c$. The shape of the potential for $\psi$ is then 
modified and new minima appear in  $\psi=\pm M$.  $\psi$ will thus roll down
into one of these new minima and, as a consequence, inflation will end. 

 As far as inflation
is concerned, hybrid inflation scenarios 
correspond effectively to single-field models with a 
potential 
characterized by $V''(\phi)>0$ and $0<\epsilon<\eta$.
A typical potential is
\beq
 V(\phi)=\Lambda^4 \left[1+\left({\phi\over\mu}\right)^p\right].
\eeq 
Once more, this potential can be seen as the lowest order in a Taylor 
expansion about the origin. 

In the case of hybrid models, the 
value $\phi_N$ of the scalar field as a function of the number of 
e-folds before the end of inflation is not determined by the above potential
and, therefore, $(\phi_N/\mu)$ can be considered as a freely adjustable 
parameter.

\end{itemize}

Many more details on inflationary models can be found in e.g. 
\cite{linde,lr,ll}.

\section{The theory of cosmological perturbations}

So far, we have concentrated our attention on strictly homogeneous and 
isotropic aspects of cosmology. Of course, this idealized version, 
although extremely useful, is not sufficient to account for real 
cosmology and it is now time to turn to the study of deviations from 
homogeneity and isotropy. 

In cosmology, inhomogeneities grow because of the attractive nature of 
gravity, which implies that inhomogeneities were much smaller in the past.
As a consequence, for most of their evolution, inhomogeneities can 
be treated as {\it linear perturbations}. The linear
treatment  ceases to be valid  on small 
scales in our recent past, hence the difficulty to reconstruct the 
primordial inhomogeneities from large-scale structure, but it is 
quite adequate to describe  the fluctuations of the CMB at the time 
of last scattering. This is  the reason why the CMB is currently 
the best observational probe of primordial inhomogeneities.
For more details on the relativistic theory of cosmological perturbations, 
which will be briefly introduced in this chapter, the reader is invited 
to read the standard reviews \cite{cosmo_perts} in the literature.

From now on, we will be mostly working with the conformal time $\eta$, 
instead of the cosmic time $t$. The conformal time is defined as 
\beq
\eta=\int {dt\over a(t)}
\eeq
so that the (spatially flat) FLRW metric takes the remarkably simple 
form
\beq
ds^2=a^2(\eta)\left[-d\eta^2+\d_{ij}dx^idx^j\right].
\label{metric_eta}
\eeq

\subsection{Perturbations of  the geometry}
Let us start with the linear perturbations of the geometry. The most general
linear perturbation of the FLRW metric can be expressed as
\beq
ds^2=a^2\left\{ -(1+2A)d\eta^2+2B_idx^id\eta+
\left(\d_{ij}+h_{ij}\right) dx^idx^j \right\} 
\label{metpert},
\eeq
where we have considered only the spatially flat FLRW metric.

We have introduced a time plus space decomposition of the perturbations.
The indices $i$, $j$ stand for {\it spatial} indices and the perturbed 
quantities   defined in (\ref{metpert}) 
can  be seen as three-dimensional tensors, for which the 
indices can be  lowered (or raised) by the spatial metric 
$\d_{ij}$ (or its inverse).

It is very convenient to separate the perturbations into three categories, 
the so-called ``scalar'', ``vector'' and ``tensor'' modes.
For example, a spatial vector field  $B^i$ can be decomposed uniquely into
a longitudinal part and a transverse part, 
\beq
B_i=\nabla_i B+\B_i, \quad \nabla_i\B^i=0,
\label{Bi}
\eeq
where the longitudinal part is curl-free and can thus be expressed as 
a gradient, and the transverse part is divergenceless. One thus gets 
one ``scalar'' mode, $B$ , and two ``vector'' modes  $\B^i$ (the index $i$ 
takes three values but the divergenceless condition implies that only 
two components are independent.

A similar procedure applies to the symmetric tensor $h_{ij}$, which can 
be decomposed as 
\beq
h_{ij}=2C\d_{ij}+2 \nabla_i\nabla_j E+2\nabla_{(i}E_{j)}
+\overline{E}_{ij}, 
\label{hij}
\eeq
with $\overline{E}^{ij}$ transverse and traceless (TT), 
i.e. $\nabla_i\overline{E}^{ij}=0$ (transverse) and
$\overline{E}^{ij}\d_{ij}=0$ (traceless), and $E_i$ transverse.
The parentheses around the indices denote symmetrization, namely
$2\nabla_{(i}E_{j)}= \nabla_{i}E_{j} + \nabla_{j}E_{i}$.
We have thus defined 
 two scalar modes, $C$ and $E$, two vector modes, $E_i$, 
and two tensor 
modes,  $\E_{ij}$.

In the following, we will be mainly interested in the metric with only
{\it scalar} perturbations, since scalar modes are the most relevant 
for cosmology and 
they  can be treated independently
of the vector and tensor modes. In matrix notation, 
the perturbed metric will thus be of the 
form
\begin{equation}
\label{scalar_pert_metric}
g_{\mu\nu} =a^2 \left[
\begin{array}{ccc}
-(1+2A) & & \nabla_iB \\ && \\
\nabla_j B & &  \left\{ (1+2C) \d_{ij} + 2\nabla_i\nabla_jE \right\}
\end{array}
\right] \,.
\end{equation}
After the description of the perturbed geometry, we turn to the perturbations
of the matter in the next subsection. 

\subsection{Perturbations of the matter}
Quite generally, the perturbed energy-momentum tensor can be written in 
the form
\begin{equation}\label{Tmunu}
T^\mu_\nu = 
\left[
\begin{array}{ccc}
-(\rho+\delta\rho) & & q_j \\ &&\\
-q^i+(\rho+P)B^i & & (P+\delta P)\delta^i_j + \pi^i_j
\end{array}
\right] \,,
\label{pert_T}
\end{equation}
where $q_i$ is the momentum density and $\pi^i_j$ is the anisotropic 
stress tensor, which is traceless, i.e. $\pi_k^k=0$. One can then 
decompose these tensors into scalar, vector and tensor components, 
as before for the metric components, so that
\beq
q_i=\nabla_i q+\bar{q}_i, \quad \nabla_i\bar{q}^i=0,
\label{qi}
\eeq
and
\beq
\pi_{ij}=\nabla_i\nabla_j \pi-{1\over3}
\delta_{ij}\nabla_k\nabla^k\pi+2\nabla_{(i}\pi_{j)}+\bar{\pi}_{ij}, \quad 
\nabla_k \pi^k=0, \quad \nabla_k\bar{\pi}^{kl}=0, \quad \bar{\pi}^k_k=0.
\label{pi_ij}
\eeq

\subsubsection{Fluid}
A widely used description for matter in cosmology is that of a fluid.
Its homogeneous part is described by the energy-momentum tensor of 
a perfect fluid, as seen earlier, while its perturbed part can be expressed
 as 
\begin{eqnarray}
&\d T_0^0=&-\d\rho, \\
&\d T_0^i=&-\left(\rho +p\right)v^i,  
\quad \d T^0_i=\left(\rho +p\right)\left(v_i+B_i\right)\\
&\d T^i_j=&\d P\d^i_j+\pi^i_j, 
\end{eqnarray}
with  $\pi_k^k=0$ as before and where $v^i$ is the three-dimensional 
fluid velocity defined by 
\beq
\d u^i={1\over a}v^i.
\eeq
It is also possible to separate  this perturbed energy-momentum
tensor into scalar, vector and tensor parts by using decompositions for 
 $v_i$ and $\pi_{ij}$ similar to (\ref{qi}) and (\ref{pi_ij}).

\subsubsection{Scalar field}
Another type  of  matter, especially useful in the context
of inflation, is a  scalar field.
The homogeneous description has already been  given earlier
and the perturbed expression for the energy-momentum tensor follows 
immediately from  (\ref{Tscalarfield}), taking into account the metric 
perturbations as well. One finds 
\begin{eqnarray}
a^2\d T_0^0&=&-a^2\d\rho=-\phi'{\d\phi}'-a^2V'\d\phi+{\phi'}^2A,\\
a^2\d T^0_i&=& a^2q_i=-\phi'\partial_i\d\phi, \\
a^2\d T^i_j&=& -\d^i_j\left(a^2V'\d\phi+{\phi'}^2A-\phi'\d\phi'\right).
\end{eqnarray}
The last equation shows that, for a scalar field, there is no anisotropic 
stress in the energy-momentum tensor.

\subsection{Gauge transformations}
It is worth noticing that there is a fundamental difference between the 
perturbations in general relativity and the perturbations in other field 
theories where the underlying spacetime is fixed. In the latter case, 
one can define the perturbation of a given field $\phi$ as
\beq
\d\phi(p)=\phi(p)-{\bar\phi}(p),
\label{pert}
\eeq
where 
${\bar\phi}$ is the unperturbed field and $p$ is any point of  the spacetime.

In the context of general relativity, spacetime is no longer a frozen 
background but must also be perturbed if  matter is perturbed. As a 
consequence, the above  definition does not make  sense
since the perturbed quantity $\phi$ lives in the perturbed spacetime 
${\cal M}$, whereas the unperturbed quantity ${\bar\phi}$ lives in 
{\it another spacetime}: the unperturbed spacetime of reference, which 
we will denote  $\bar{\cal M}$. In order to use a definition similar 
to (\ref{pert}), one must introduce a one-to-one 
 {\it identification}, $\iota$, 
between the points of $\bar{\cal M}$ and those of ${\cal M}$.
The perturbation of the field can then be defined as 
\beq
\d\phi(\bar{p})=\phi(\iota(\bar{p}))-{\bar\phi}(\bar{p}),
\eeq
where $\bar{p}$ is a point of $\bar{\cal M}$.

However, the identification $\iota$ is not uniquely defined, and therefore 
the definition of the perturbation depends on the particular 
choice for $\iota$: two different identifications, $\iota_1$ and
$\iota_2$ say, lead to two different definitions for the perturbations.
This ambiguity can be related to the freedom of choice of the 
coordinate system. Indeed, if one is given 
a coordinate system in $\bar{\cal M}$, one 
can transport it into ${\cal M}$ via the identification. $\iota_1$ and 
$\iota_2$ thus define two different coordinate systems in ${\cal M}$, and 
in this respect, a {\it change of identification} can be seen as 
a {\it change of coordinates} in ${\cal M}$.

The metric perturbations, introduced  in  (\ref{metpert}), are 
modified in a coordinate transformation  of the form
\beq x^\alpha\rightarrow x^\alpha+\xi^\alpha, \qquad \xi^\alpha=(\xi^0,\xi^i) 
\label{transjauge}.
\eeq
It can be shown that the change of the metric components can be expressed 
as 
\beq
\Delta \left(\delta g_{\mu\nu}\right)=
-2\D_{(\mu}\xi_{\nu)}.
\eeq
where $\Delta$ represents the variation, due the coordinate transformation, 
at the same old and new coordinates (and thus at different physical points).
The above variation can be decomposed into 
individual variations  for the various components of the metric defined 
earlier. One finds
\begin{eqnarray}
\Delta A&=&-{\xi^0}'-\h\xi^0 \label{gt1}\\
           \Delta B_i&=&\nabla_i\xi^0-\xi_i' \\
           \Delta h_{ij}&=&- 2 \left(\nabla_{(i}\xi_{j)}
-\h\xi^0\d_{ij}\right). \label{gt3}
\end{eqnarray}

The effect of a coordinate transformation can also be decomposed along
the scalar, vector and tensor sectors introduced earlier.
The generator $\xi^\alpha$ of the coordinate transformation can be 
written as 
\beq  
\xi^\alpha=(\xi^0,\nabla^i\xi+\overline{\xi}^i), 
\eeq
with $\overline{\xi}^i$ transverse.
This  shows explicitly that $\xi^\alpha$ contains two scalar 
components,  $\xi^0$ and $\xi$, and two vector components,
$\overline{\xi}^i$. The transformations (\ref{gt1}-\ref{gt3}) are then 
decomposed into :
\begin{eqnarray}
 A&\rightarrow &A-{\xi^0}'-\h\xi^0 \nonumber\\
 B&\rightarrow &B+\xi^0-\xi' \nonumber\\
C&\rightarrow &C-\h\xi^0  \nonumber\\
 E&\rightarrow &E-\xi \label{transfjauge} \\
\overline{B}^i&\rightarrow &\overline{B}^i-{{\overline{\xi}}^i}' \nonumber\\
 E^i&\rightarrow &E^i-\overline{\xi}^i. \nonumber
 \end{eqnarray}
The tensor perturbations remain unchanged since $\xi^\alpha$ does not
contain any tensor component.
The matter perturbations, either in the fluid description or in the 
scalar field description, follow similar transformation laws in 
a coordinate change.

In order to study the physically relevant modes and not spurious modes
due to coordinate ambiguities, two strategies can a priori be envisaged.
The first consists in working from the start in a specific gauge. 
A familiar choice in the literature on cosmological perturbations is 
the {\it longitudinal gauge} (also called conformal Newton gauge), which 
imposes 
\beq
B_\L=0, \quad E_\L=0.
\eeq

The second approach consists in defining {\it gauge-invariant variables}, 
i.e. variables that are left unchanged under a coordinate transformation.
For the scalar metric perturbations, we start with four quantities ($A$, $B$,
$C$ and $E$) and we can use two gauge transformations ($\xi^0$ and $\xi$).
This implies that the scalar metric perturbations must be 
 described by {\it two} independent gauge-invariant quantities. 
Two such quantities are 
\beq\Phi=  A+(B-E')'+\h (B-E')    \eeq
and
\beq\Psi=- C -\h (B-E'),    \eeq
as it can be checked by considering the explicit transformations in 
(\ref{transfjauge}). 
It turns out that, in the longitudinal gauge, the remaining scalar 
perturbations $A_\L$ and $C_\L$ are numerically 
equivalent  to the gauge-invariant 
quantities just defined $\Phi$ and $-\Psi$. 

In practice, one can combine  the two strategies by doing explicit
calculations in a given gauge and then by relating the quantities defined 
in this gauge to some gauge-invariant variables. It is then possible 
to translate the results in any other gauge.
In the rest of these lectures, we will use the longitudinal gauge.

\subsection{The perturbed Einstein equations}
After having defined the metric  and the matter perturbations, 
we can now relate them via  the perturbed Einstein 
equations. 
We will consider here explicitly {\it only the 
scalar sector}, which is the most complicated but also the most interesting 
for cosmological applications.

Starting from the perturbed metric (\ref{scalar_pert_metric}), 
one can compute the 
components of Einstein's tensor at linear order. In the {\it longitudinal 
gauge}, i.e. with  $B_\L=E_\L=0$, one finds
\begin{eqnarray}
\label{dG00}
\left(\d G^0_0\right)_\L&=&
{2\over a^2}\left[3\h^2A_\L-3\h C'_\L+\nabla^2 C_\L\right] \\
\label{dG0i}
\left(\d G^0_i\right)_\L&=&
{2\over a^2}\nabla_i\left[-\h A_\L+C'_\L\right] \\
\label{dGij}
\left(\d G^i_j\right)_\L&=&
{1\over a^2}\nabla^i\nabla_j\left(-C_\L-A_\L\right) 
+{1\over a^2}\left[-2 C_\L''-4\h C_\L'+\nabla^2 C_\L \right. \cr
& & \left.+2\h A_\L'
+\nabla^2 A_\L+2\left(2\h'+\h^2\right)A_\L
\right]\delta^i_j.
\end{eqnarray}

Combining with the perturbations of the energy-momentum tensor given 
in (\ref{pert_T}), the 
perturbed Einstein equations yield, {\it in the longitudinal gauge}, the 
following relations: the energy constraint (from (\ref{dG00}))
\beq
3\h^2 A_\L-3\h C_\L'+\nabla^2 C_\L=-4\pi G a^2\d\rho_\L,
\label{energy}
\eeq
 the momentum constraint (from (\ref{dG0i}))
\beq
C_\L'-\h A_\L=4\pi G\,  a^2\,  q_\L,
\label{momentum}
\eeq
the ``anisotropy constraint'' (from the traceless part of (\ref{dGij}))
\beq
-A_\L-C_\L=8\pi G \, a^2\, \pi_\L,
\label{anisotropy}
\eeq
and finally 
\beq
C_\L''+2\h C_\L'- \h A_\L'-(2\h'+\h^2)A_\L-{1\over 3}\nabla^2(A_\L+C_\L)
=-4\pi G\, \d P_\L,
\label{dynamics}
\eeq
obtained from the trace of (\ref{dGij}).

The  combination of the energy and 
momentum constraints gives the useful relation 
\beq
\nabla^2\Psi= 4\pi G a^2\left(\d\rho_\L- 3\h q_\L\right)\equiv 4\pi G a^2
\d\rho_c,
\label{poisson}
\eeq
where we have introduced the {\it comoving} energy density perturbation
 $\d\rho_c$:  this gauge-invariant quantity corresponds, according to its 
definition,  to the  energy density perturbation measured  
 in comoving gauges characterized by $\d T^0_i=q_i=0$. We have 
also replaced $C_\L$ by $-\Psi$. Note that 
the above equation is quite similar to the Newtonian Poisson equation, but 
with quantities whose natural interpretation is given in {\it different} 
gauges.

\subsection{Equations for the matter}
As mentioned earlier, a consequence of Einstein's equations is that the 
{\it total} energy-momentum tensor is covariantly conserved (see 
Eq.~(\ref{DT0})). For a fluid, 
the conservation of the energy-momentum tensor leads to a continuity
 equation that generalizes  the {\it continuity equation} of fluid 
mechanics, and a momentum conservation equation that generalizes   the 
{\it Euler equation}. In the case of a single fluid, combinations of 
the perturbed Einstein equations  obtained in the previous 
subsection  lead necessarily 
to the perturbed continuity and Euler equations for the fluid. 
In the case of several non-interacting fluids, however, one must 
impose {\it separately} the covariant conservation of {\it each} 
energy-momentum tensor: this is not a consequence of Einstein's equations,
which impose only the conservation of the {\it total} energy-momentum
tensor. 

In the perspective to deal with several cosmological fluids, it is therefore
useful to write the perturbation equations, satisfied by a given 
fluid, that follow 
 only from the conservation of the corresponding 
energy-momentum tensor, and 
independently of Einstein's equations.

The continuity equation can be obtained by perturbing 
$u^\mu\D_\nu T^\nu_\mu=0$. One finds, in  {\it any gauge},
\beq
\d\rho'+3\h \left(\d\rho+\d P\right)+\left(\rho+P\right)\left(3 C'
+ \nabla^2 E'+\nabla^2v\right)=0.
\eeq
Dividing by $\rho$, this can be reexpressed in terms of the density contrast
$\d=\d\rho/\rho$:
\beq
\d'+3\h \left({\d P\over \d\rho}-w\right) \d 
+(1+w)\left(\nabla^2 v +3C'+\nabla^2 E'\right)=0,
\label{conserv_pert}
\eeq
where $w=p/\rho$ ($w$ is not necessarily constant here).
The perturbed Euler equation is derived from the spatial projection 
of $\d(\D_\nu T^\nu_\mu)=0$.
This gives
\beq
\left(v+B\right)' +\left(1-3c_s^2\right)\h \left(v+B\right)
+A+{\d P\over \rho+ P}+{2\over 3(\rho+P)}
\nabla^2\pi=0,
\label{euler_pert}
\eeq
where $c_s$ is the sound speed, which is related to the time derivatives 
of the background energy density and pressure:
\beq
c_s^2={p'\over\rho '}.
\eeq
There are as many systems of equations (\ref{conserv_pert}-\ref{euler_pert})
as the number of fluids. If the fluids are interacting, one must add 
an interaction term on the right-hand side of the Euler equations.

Finally, let us stress that the fluid description is not always an adequate
approximation for cosmological matter. A typical example is the  
photons during and after recombination: their mean free path becomes so large
that they must be treated as a gas, which requires the  use of the  Boltzmann 
equation (see e.g. \cite{ma_bertschinger} for a presentation of  the 
Boltzmann equation in the cosmological context).

\subsection{Initial conditions for standard cosmology}
The notion of {\it initial conditions} depends in general on the 
context, since 
the initial conditions for a given period in the history of the 
universe can be seen as the final conditions of the previous phase.
In  cosmology,  ``initial conditions'' usually refer to the state 
of the perturbations  during 
 the {\it radiation dominated era} (of standard cosmology) 
 and on   {\it wavelengths larger 
 than the Hubble radius}. 

Let us first address in details  the question of 
initial conditions in the simple case  of a single 
perfect fluid, radiation,   
with equation of state $p=\rho/3$ (which gives $c_s^2=w=1/3$).
The four key equations are the continuity, Euler, Poisson and anisotropy 
equations, respectively Eqs~(\ref{conserv_pert}), (\ref{euler_pert}), 
(\ref{poisson}) and (\ref{anisotropy}). In terms of the Fourier components,
\beq
Q(\k)=\int {d^3 x\over (2\pi)^{3/2}}e^{-i \k.\x} Q(\x),
\eeq
and of the dimensionless quantity
\beq
x\equiv k\eta
\eeq
(during the radiation dominated era $\h=1/\eta$), the four equations 
can be rewritten as 
\begin{eqnarray}
&& {d\d\over dx}-{4\over 3}\V+4{dC\over dx}=0,\label{ic1}\\
 && {d\V\over dx  }+{1\over 4}\d +A=0, \label{ic2}\\
&& x^2 C={3\over 2}\left(\d -{4\over x}\V\right)\label{ic3}\\
&& C=-A.
\label{ic4}
\end{eqnarray}
We have introduced the quantity 
 $\V\equiv kv$, which has the dimension of a velocity.
Since we are interested in perturbations with wavelength larger than 
the Hubble radius, i.e. such that $x=k/\h \ll 1$, 
it is useful to consider a Taylor expansion for the 
various perturbations, for instance  
\beq
\V=\V^{(0)}+x\V^{(1)}+{x^2\over 2}\V^{(2)}+\dots
\eeq
One then substitutes these Taylor expansions into the above 
system of equations. In particular,   the Poisson equation (\ref{ic3})
gives 
\beq
\V^{(0)}=0,
\eeq 
in order to avoid a divergence, as well as  
\beq
\V^{(1)}={1\over 4}\d^{(0)}. 
\eeq
The Euler equation (\ref{ic2}) then  gives 
\beq
A^{(0)}=-{1\over 2}\d^{(0)}.
\eeq 
The conclusion is that the initial conditions for each Fourier 
mode are determined  by a single quantity, e.g. 
$\d^{(0)}$,  the other quantities being related  via 
the above constraints.

In general, one must consider several cosmological fluids. 
Typically, the ``initial'' or ``primordial'' perturbations
are  defined deep in the radiation era but at temperatures low enough, 
i.e. after nucleosynthesis, so that the main cosmological 
components reduce to  the usual  photons, baryons,  neutrinos and 
cold dark matter (CDM).
The system 
(\ref{ic1}-\ref{ic4}) must thus be generalized to include  a continuity
equation and a Euler equation for each fluid. 
The above various cosmological 
species can be characterized by their number density,
 $n_X$, and their energy density $\rho_X$. 
In a multi-fluid system, it is useful to distinguish {\it adiabatic} and
{\it isocurvature} perturbations. 

 The {\it adiabatic mode} is defined as a 
perturbation affecting all the cosmological  species such that the relative 
 ratios in the number densities remain unperturbed, i.e. such that
\beq
\delta\left(n_X/n_Y\right) =0.
\label{nullratio}
\eeq
It is associated with a curvature perturbation, via Einstein's 
equations,  since there is  a global perturbation of the matter content.
This is why the adiabatic perturbation
 is also called {\it curvature} perturbation. 
In terms of the energy density contrasts, 
the adiabatic perturbation is characterized by the relations 
\beq
{1\over 4}\d_\gamma={1\over 4}\d_\nu={1\over 3}\d_b={1\over 3}
\d_c,
\eeq
They follow directly from the prescription (\ref{nullratio}), each 
coefficient depending on the equation of state of the particuler species.

 \def\dc{{\delta}}
Since there are several cosmological species, it is also possible to perturb 
the matter components without perturbing the geometry. This corresponds
to {\it isocurvature} perturbations,  characterized by 
variations in the particle number ratios but with vanishing curvature 
perturbation.
The variation in the relative particle number densities between 
two species can be quantified by the so-called {\it entropy perturbation}
\beq
S_{A,B}\equiv {\dc n_A\over n_A}-{\dc n_B\over n_B}.
\eeq
 When the equation of state for a given species is such that
$w\equiv p/\rho= {\rm Const}$, then one can reexpress the entropy
perturbation in terms of the density contrast, in the form
\beq
S_{A,B}\equiv {\d_A\over 1+w_A}-{\d_B\over 1+w_B}.
\eeq
It is convenient to choose a species of reference, for instance the 
photons, and to define the entropy perturbations of the other species 
relative to it:
\begin{eqnarray}
S_b&\equiv \d_b-{3\over 4} \d_\gamma, \\
S_c&\equiv \d_c-{3\over 4} \d_\gamma, \\
S_\nu&\equiv {3\over 4}\d_\nu-{3\over 4} \d_\gamma,
\end{eqnarray}
thus  define respectively the {\it baryon isocurvature mode}, 
the {\it CDM isocurvature mode}, the {\it neutrino isocurvature mode}.
In terms of the entropy perturbations, the adiabatic mode is 
obviously characterized by $S_b=S_c=S_\nu=0$.

In summary, we can decompose a general perturbation, described by 
four density contrasts, into one adiabatic mode and three isocurvature 
mode. In fact, the problem is slightly more complicated because the evolution
of the initial  velocity fields. For a single fluid, we have seen that 
the velocity field is not an independent initial condition but depends
on the density contrast so that there is no divergence backwards in time.  
In the case of the four species mentioned above, 
 there remains however one arbitrary relative velocity 
between the species, which gives an additional mode, usually named 
the {\it neutrino isocurvature  velocity} perturbation.

The CMB is a powerful way to study isocurvature perturbations because  
 (primordial) adiabatic and isocurvature perturbations 
produce very distinctive 
features in the CMB anisotropies \cite{smoot}. 
Whereas an adiabatic initial perturbation  generates a cosine 
oscillatory mode in the photon-baryon fluid, leading to 
an acoustic peak at $\ell\simeq 220$ (for a flat universe), a
  pure isocurvature initial perturbation generates a sine oscillatory 
mode resulting in a first peak at $\ell \simeq 330$. 
The unambiguous observation of the first peak at $\ell\simeq 220$ has 
eliminated the possibility of a dominant isocurvature perturbation. 
The recent observation by WMAP of the CMB polarization has also confirmed 
that the initial perturbation is mainly an adiabatic mode. But this does not 
exclude the presence of a subdominant isocurvature contribution, which 
could be detected in  future high-precision experiments such as Planck.

\subsection{Super-Hubble evolution}
In the case of {\it adiabatic perturbations}, there is only one 
(scalar) dynamical degree of freedom. 
One can thus choose   either an energy density perturbation or
  a  metric perturbation, to study the dynamics, 
the other quantities being determined by the constraints.

If one considers the metric perturbation $\Psi=\Phi$ (assuming $\pi=0$), 
one can  
combine Einstein's equations (\ref{dynamics}), with $\d p=c_s^2\d\rho$
(for adiabatic perturbations), and (\ref{energy}) to obtain  
a second-order differential 
equation in terms of $\Psi$ only:
\beq
\Psi''+3\h (1+c_s^2)\Psi'+\left[2\h'+(1+3c_s^2)\h\right]\Psi- k^2c_s^2 \Psi=0.
\eeq
Using the background Friedmann equations,  the sound speed can be reexpressed 
in terms of the scale factor and its derivatives.
For scales larger than the sonic horizon, i.e. such that 
 $kc_s\ll \h$, the above equation can be integrated explicitly and yields
\beq
\Psi={\h\over a^2}\left[\alpha \int d\eta {a^2\left(\h'-\h^2\right)\over \h^2}+\beta\right], 
\eeq
where $\alpha$ and $\beta$ are two integration constants.

For a scale factor evolving like 
 $a\propto t^p$, one gets 
\beq
\Psi=-{\alpha\over p+1}+\beta\,  p\,  t^{-p-1}.
\eeq
They are two modes: a constant mode and a decaying mode.
Note that, in the previous subsection on the initial conditions, 
we eliminated the decaying mode to avoid the divergence when going 
backwards in time. 
 
In a transition between two cosmological phases characterized respectively 
by the scale factors
 $ a\propto t^{p_1}$ and  $a\propto t^{p_2}$, 
one can easily finds the relation between the asymptotic behaviours of 
$\Psi$ (i.e. after the decaying mode becomes negligible) by using the 
constancy of  $\alpha$.
This gives  
\beq
\Psi_2={p_2+1\over p_1+1}\Psi_1.
\eeq
This is valid  only asymptotically.
In the case of a sharp transition, 
 $\Psi$ must be continuous at the transition and the above relation will apply
only after some relaxation time.
For a transition radiation/non-relativistic matter, one finds 
\beq
\label{rad_matt}
\Psi_{mat}={9\over 10}\Psi_{rad}.
\eeq

In practice and for more general cases, it turns out that it is much 
more convenient to follow the evolution 
 of cosmological perturbations by resorting  to quantities that 
 are {\it conserved on super-Hubble scales}.
A familiar  example of such a quantity is the {\it curvature perturbation 
on uniform-energy-density hypersurfaces}, which can be expressed 
in any gauge  as 
\beq
\zeta=C-\h {\delta\rho\over  \rho'}.
\label{zeta}
\eeq
This is a gauge-invariant quantity by definition. 
The conservation equation (\ref{conserv_pert}) can then be rewritten as 
\beq
\zeta'=-{\h\over \rho + P}\d P_{nad}-{1\over 3}\nabla^2(E'+v),
\label{zetaprim}
\eeq
where $\d P_{nad}$ is the non-adiabatic part of the pressure perturbation, 
defined by
\beq
\d P_{nad}=\d P-c_s^2\d\rho.
\eeq
The expression (\ref{zetaprim}) shows that $\zeta$ is conserved 
{\it on super-Hubble scales} in the case of {\it adiabatic perturbations}.

Another convenient quantity, 
which is sometimes used in the literature instead of 
$\zeta$, is the {\it curvature perturbation on comoving hypersurfaces}, 
which can be written in any gauge as 
\beq
-\R=C+{\h\over \rho+P} q.
\label{R}
\eeq
It is easy to relate the two quantities $\zeta$ and $\R$. Substituting 
e.g. $\d\rho=\d\rho_c+ 3\h q$, which follows from 
the definition (see (\ref{poisson})) of  the comoving energy 
density perturbation, into (\ref{zeta}), one finds 
\beq
\zeta=-\R+{\d\rho_c\over \rho+P}.
\eeq
Using Einstein's equations, in particular (\ref{poisson}), 
this can be rewritten as 
\beq
\zeta=-\R-{2\rho\over 3(\rho+P)}\left({k\over aH}\right)^2\Psi.
\eeq
The latter expression shows that $\zeta$ and $\R$ coincide in the 
super-Hubble limit $k\ll aH$.

The quantity $\R$  can also 
 be expressed in terms of the two Bardeen potentials 
$\Phi$ and $\Psi$. Using the momentum constraint (\ref{momentum}) and 
the Friedmann equations, one finds 
\beq
\label{R_Psi}
\R=\Psi-{H\over \dot H}\left(\dot\Psi+H\Phi\right).
\eeq
In a cosmological phase dominated by a fluid 
with no anisotropic stress, so that $\Phi=\Psi$, and 
 with an equation of state $P=w\rho$ with $w$ constant, 
we have already seen that  $\Psi$ is constant with time. Since 
the scale factor evolves like $a\propto t^p$ with $p=2/3(1+w)$, 
the relation (\ref{R_Psi}) between $\R$ and $\Psi$ reduces to
\beq
\label{R_Psi2}
\R={5+3w\over 3(1+w)}\Psi.
\eeq
In the radiation era, $\R=(3/2)\Psi$, whereas in the matter era, 
$\R=(5/3)\Psi$, and since $\R$ is conserved, 
one recovers the conclusion given in Eq.~(\ref{rad_matt}).

During inflation, $w\simeq -1$ and 
\beq
w+1={\dot\phi^2\over\rho}\simeq -{2\over 3}{\dot H\over H^2}
\eeq
so  that 
\beq
\R\simeq -{H^2\over \dot H}\Psi_{inf}.
\eeq

For a scalar field, the perturbed equation of motion reads
\beq
\ddot{\d\sigma}+3H\dot{\d\sigma}+\left({k^2\over a^2}+V''\right)
\d\sigma
=\dot\sigma \left(\dot\Phi+3\dot \Psi\right)-2V' \Phi.
\eeq

\section{Quantum fluctuations and ``birth'' of cosmological perturbations}
In the previous section, we have discussed the  {\it classical} 
evolution of the cosmological perturbations. In the {\it classical} context,
the initial conditions, defined deep in the radiation era, 
are a priori arbitrary. What is remarkable with inflation is that 
the accelerated expansion can convert  {\it initial 
vacuum quantum fluctuations} into ``macroscopic'' cosmological perturbations
(see \cite{quantum} for the seminal works). In this sense, inflation 
 provides us with ``natural'' initial conditions, which turn out to 
be the initial conditions that agree with the present observations. 

\subsection{Massless scalar field in  de Sitter}
As a warming-up, it is instructive to  discuss
 the case of a massless scalar field
in a so-called de Sitter universe, or a FLRW spacetime with exponential 
expansion, $a\propto \exp(Ht)$. In conformal time, the scale 
factor is given by
\beq
a(\eta)=-{1\over H\eta}.
\eeq
The conformal time is here negative (so that the scale factor is positive) and 
goes from $-\infty$ to $0$.
The action for a massless scalar field in this geometry  is given by 
\beq
\label{action_dS}
S=\int d^4x \sqrt{-g}\left(-{1\over 2}\partial_\mu\phi \partial^\mu\phi\right)
 =\int d\eta \, \,  d^3x\, \, 
 a^4\left[{1\over 2 a^2}{\phi'}^2-{1\over 2 a^2}{\vec\nabla \phi}^2\right],
\eeq
where we have substituted in the action the cosmological metric  
(\ref{metric_eta}). 
Note that, whereas we still allow for spatial
variations of the scalar field, i.e. inhomogeneities, we will assume here, 
somewhat  inconsistently, that the geometry is completely fixed as 
homogeneous. We will deal later with the question of  the metric perturbations.

 It is possible to write the above action with a canonical kinetic term
via the change of variable 
\beq
u=a\phi.
\eeq
After an integration by parts, the action (\ref{action_dS}) 
 can be rewritten as  
\beq
S={1\over 2}\int d\eta \, \, d^3x \,  
\left[{u'}^2-  {\vec\nabla u}^2 +{a''\over a}u^2\right].
\eeq
The first two terms  are familiar since they are the same as 
in 
the action for a free massless scalar field in Minkowski spacetime.
The fact that our scalar field here lives in de Sitter spacetime rather 
than Minkowski has been reexpressed as a {\it time-dependent 
effective mass} 
\beq
m^2_{eff}=-{a''\over a}=-{2\over \eta^2}.
\eeq

Our next step will be to quantize the scalar field $u$ by using 
the standard procedure of quantum field theory.
One first turns $u$ into a quantum field denoted $\hat u$, which we 
expand in Fourier space as 
\beq
\label{Fourier_quantum}
\hat u (\eta, \vec x)={1\over (2\pi)^{3/2}}\int d^3k \left\{{\hat a}_{\vec k} u_k(\eta) e^{i \vec k.\vec x}
+ {\hat a}_{\vec k}^\dagger u_k^*(\eta) e^{-i \vec k.\vec x} \right\},
\eeq
where  the $\hat a^\dagger$ and  $\hat a$ are 
 creation and annihilation operators, which satisfy the 
usual commutation rules 
\beq
\label{a}
\left[ {\hat a}_{\vec k}, {\hat a}_{\vec k'}\right]= 
\left[ {\hat a^\dagger}_{\vec k}, {\hat a^\dagger}_{\vec k'}\right]= 0, 
\quad
\left[ {\hat a}_{\vec k}, {\hat a^\dagger}_{\vec k'}\right]= \delta(\vec k-\vec k').
\eeq
The function $u_k(\eta)$ is a complex time-dependent function that  
must satisfy the {\it classical} equation of motion in Fourier space, namely
\beq
\label{eom_u}
u_k''+\left(k^2-{a''\over a}\right)u_k=0, 
\eeq
which is simply the equation of motion for an oscillator with 
a time-dependent mass. In the case of a massless scalar field in 
Minkowski spacetime, this effective mass is zero ($a''/a=0$) 
and one usually takes $u_k=(\hbar/ 2k)^{1/2}e^{-ik\eta}$ (the choice for the 
normalization factor will be clear below).
In the case of de Sitter, one can  solve explicitly the above 
equation with $a''/a=2/\eta^2$ and the general solution is given by
\beq
\label{general_solution}
u_k=\alpha e^{-ik\eta }\left(1-{i\over k\eta}\right)
+\beta e^{ik\eta }\left(1+{i\over k\eta}\right).
\eeq

Canonical quantization  consists in imposing the following commutation rules
on the $\eta=$constant hypersurfaces:
\beq
\left[\hat u(\eta, \vec x), \hat u(\eta, \vec x')\right]=
 \left[\hat\pi_u(\eta, \vec x), \hat\pi_u(\eta, \vec x')\right]=0
\eeq
and 
\beq
\label{canonical}
\left[\hat u(\eta, \vec x), \hat\pi_u(\eta, \vec x')\right]=i\hbar \delta(\vec x-\vec x'), 
\eeq
where $\pi_u\equiv \delta S/\delta u'$ is the conjugate momentum of $u$.
In the present case, 
 $\pi_u=u'$ since the kinetic term is canonical.

Subtituting the expansion (\ref{Fourier_quantum}) in 
the commutator (\ref{canonical}), and using the commutation rules 
for the creation and annihilation operators (\ref{a}), 
one obtains the relation
\beq
u_k {u'_k}^*-u_k^*u'_k=i\hbar,
\label{wronskien}
\eeq
which determines the normalization of the Wronskian.

The choice of a specific  function $u_k(\eta)$  
corresponds to a particular 
 prescription for the physical vacuum $| 0 \rangle$, 
defined by
\beq
{\hat a}_{\vec k}|0\rangle=0.
\eeq 
A different choice for $u_k(\eta)$ is associated to a different decomposition
into creation and annihilitation modes and thus to a different vacuum. 

Let us now  note that the wavelength associated with 
a given mode $k$ can always be found {\it within} the Hubble radius 
provided one goes sufficiently far  backwards  in time, since 
the comoving Hubble radius is shrinking during inflation. 
In other words, for $|\eta|$ sufficiently big,
one gets $k|\eta|\gg 1$. 
Moreover, for a wavelength smaller than the Hubble radius, 
one can neglect the influence of the curvature of spacetime and the 
mode behaves as in a Minkowski spacetime, as can also be checked explicitly 
with the equation of motion (\ref{eom_u}) 
(the effective mass is negligible for 
$k|\eta|\gg 1$).
Therefore, the most natural physical prescription is to take  the particular
solution that corresponds to the usual Minkowski vacuum, i.e. 
$u_k\sim \exp(-ik\eta)$, in the limit $k|\eta|\gg 1$.
In view of (\ref{general_solution}), this corresponds to the choice
\beq
u_k=\sqrt{\hbar\over 2k}e^{-ik\eta }\left(1-{i\over k\eta}\right), 
\label{u_k}
\eeq 
where the coefficient has been determined by the normalisation condition
(\ref{wronskien}). 
This choice, in the jargon of quantum field theory on curved 
spacetimes, corresponds to the {\it Bunch-Davies vacuum}.

Finally, one can compute the {\it correlation function} for the  
scalar field $\phi$ in the vacuum state defined above. When Fourier 
transformed, the correlation function defines the {\it power 
spectrum} $\P_\phi(k)$:
\beq
\langle 0| \hat \phi(\vec x_1) \hat\phi(\vec x_2)|0\rangle= 
\int d^3 k \, \, e^{i \vec k.(\vec x_1-\vec x_2)}
{\P_\phi(k)\over 4\pi k^3}.
\eeq
Note that the homogeneity and isotropy of the quantum field 
is used implicitly in the definition of the power spectrum, which is 
``diagonal'' in Fourier space (homogeneity) and depends only on the 
norm of  $\vec k$ (isotropy).
In our case, we find 
\beq
\label{ps_phi}
2\pi^2k^{-3}\P_\phi= {|u_k|^2\over a^2},
\eeq
which gives in the limit when the wavelength is {\it larger than  the 
Hubble radius}, i.e. $k|\eta|\ll 1$, 
\beq
\P_\phi(k)\simeq \hbar \left({H\over 2\pi}\right)^2 
\qquad (k\ll aH)
\eeq
Note that, in  the opposite limit, 
i.e. for wavelengths smaller than the Hubble radius ($k|\eta|\gg 1$), 
one recovers the usual result for fluctuations in 
Minkowski vacuum, $\P_\phi(k)=\hbar (k/2\pi a)^2$.

We have used  a quantum description of the scalar field. 
But the cosmological 
perturbations are usually  described  by {\it classical random 
fields}. 
Roughly speaking, the transition between the quantum and classical (although
stochastic) descriptions makes sense  when the perturbations  exit 
 the Hubble radius.  Indeed
 each of the terms in the 
 Wronskian (\ref{wronskien}) is roughly of the order 
 $\hbar/2(k\eta)^3$ in the super-Hubble limit and the non-commutativity 
can then be neglected. In this sense, one can see the exit outside   the 
Hubble radius as a quantum-classical transition, although much refinement 
would be  needed to make this statement more precise. 

\subsection{Quantum fluctuations with metric perturbations}
Let us now move to the more realistic case of a perturbed inflaton field  
living in a perturbed cosmological geometry. 
The situation is more complicated than in the previous  problem, 
because Einstein's equations imply 
that scalar field fluctuations must  necessarily coexist  with {\it metric 
fluctuations}. 
A correct treatment, either classical or quantum, must thus involve 
both the scalar field perturbations and metric perturbations.

In order to quantize this coupled 
system, the easiest procedure consists in 
identifying the {\it true degrees of freedom}, the other variables being 
then derived from them via  constraint equations. As we saw in the 
classical analysis of cosmological perturbations, there exists only one 
scalar degree of freedom in the case of a single scalar field, which we must 
now identify.

 The starting point is the action of the coupled system scalar field 
plus gravity expanded up to second order in the linear perturbations.
Formally this can be written as 
\beq
S[\bar \phi+\d\phi, g_{\mu\nu}=\bar g_{\mu\nu}+h_{\mu\nu}]= S^{(0)}[\bar\phi, \bar g_{\mu\nu}]
+S^{(1)}[\d\phi, h_{\mu\nu};\bar\phi, \bar g_{\mu\nu}]
+S^{(2)}[\d\phi, h_{\mu\nu};\bar\phi, \bar g_{\mu\nu}],
\eeq
where the first term
 $S^{(0)}$ contains only the homogeneous part, 
 $S^{(1)}$ contains all terms linear in the perturbations (with coefficients
depending on the homogeneous variables), and finally 
 $S^{(2)}$ contains the terms quadratic in the linear perturbations.
When one substitutes the FLRW equations of motion in 
$S^{(1)}$ (after integration by parts), 
one finds that $S^{(1)}$ vanishes, which is not very surprising since 
this is how one gets the homogeneous equations of motion, via 
the Euler-Lagrange equations,
from the variation of the action.

The term  $S^{(2)}$ is the piece we are interested in: the corresponding
Euler-Lagrange equations give the equations of motion  for the linear 
perturbations, which we have already obtained; but more importantly, this 
term enables us to quantize the linear perturbations and to find the 
correct normalization. 

If one restricts oneself to the scalar sector, the quadratic part of the 
action depends on the four metric perturbations
 $A$, $B$, $C$, $E$,  as well as on
the scalar field perturbation  $\d\phi$. After some cumbersome manipulations,
by using the FLRW equations of motion, one can show 
that the second-order action for scalar perturbations can be rewritten 
in terms of a single variable \cite{Mukhanov:rq}
\beq 
\label{v}
v=a\left(\d\phi-{\phi'\over \h}C\right),
\eeq
which is a linear combination mixing scalar field {\it and} metric 
perturbations. The variable 
$v$ represents the true dynamical degree of freedom of 
the system, and one can check immediately that it is indeed a 
gauge-invariant variable.

In fact, $v$ is proportional to the comoving curvature perturbation defined 
in (\ref{R}) and which, in the case of a single scalar 
field, takes the form
\beq
\R=-C+{\h\over \phi'}\delta\phi.
\eeq
Note also that, if one can check a posteriori that $v$ is the variable 
describing the true degree of freedom by expressing the action in terms 
of $v$ only (modulo, of course, a multiplicative factor depending only 
on homogeneous quantities: the $v$ defined here is such that it gives 
a canonical kinetic term in the action), one can identify $v$ in a 
systematic way by resorting to Hamiltonian techniques, in particular 
the Hamilton-Jacobi equation \cite{l94}.

With the variable $v$, the quadratic action takes the extremely simple form
\beq
S_v={1\over 2}\int d\eta \, \, d^3x\, \left[{v'}^2+\partial_i v \partial ^i v
+{z''\over z}v^2\right],
\eeq
with
\beq
\label{z}
z=a{\phi'\over \h}.
\eeq
This action is thus 
analogous to that of a scalar field in Minkowski spacetime 
with a time-dependent mass. One is thus back in a situation similar 
to the previous subsection, with the notable difference that the 
effective time-dependent mass is now
$z''/z$,  instead of $a''/a$.  

The quantity we will be eventually interested in is the comoving 
curvature perturbation $\R$, which is related to the canonical variable 
$v$ by the relation 
\beq
v=-z\R.
\eeq
Since, by analogy with (\ref{ps_phi}),
 the power spectrum for $v$ is given by
\beq
2\pi^2 k^{-3}{\cal P}_v(k)= |v_k|^2, 
\eeq
the corresponding power spectrum for $\R$ is found to be 
\beq
\label{ps_R}
2\pi^2 k^{-3}{\cal P}_\R(k)= {|v_k|^2\over z^2}.
\eeq

In the case of an inflationary phase in the {\it slow-roll} approximation, 
 the evolution of 
 $\phi$ and of  $H$ is much slower than 
 that of the scale factor $a$. 
Consequently, one gets approximately 
\beq
{z''\over z}\simeq {a''\over a},  \qquad{\rm (slow-roll)}
\eeq
and all results of the previous section obtained for  $u$
apply directly  to our variable $v$ in the slow-roll approximation.
This implies that the properly normalized function corresponding to 
the Bunch-Davies vacuum is approximately given by 
\beq
v_k\simeq \sqrt{\hbar\over 2k}e^{-ik\eta }\left(1-{i\over k\eta}\right). 
\label{v_k}
\eeq 
In the super-Hubble limit $k|\eta| \ll 1$ the function $v_k$ behaves 
like
\beq
v_k\simeq - \sqrt{\hbar\over 2k}\, {i\over k\eta} \simeq 
i\sqrt{\hbar\over 2k}\, {aH\over k}, 
\eeq 
where we have used $a\simeq - 1/(H\eta)$.
Consequently, on scales larger than the Hubble radius, 
 the power spectrum for $\R$  is 
found, combining (\ref{ps_R}), (\ref{z}) and (\ref{v_k}),
  to be given by 
\beq
\P_\R\simeq {\hbar\over 4\pi^2}\left({H^4\over \dot\phi^2}\right)_{k=aH},
\label{power_S}
\eeq
where we have reintroduced the cosmic time instead of the conformal time. 
This is the famous result for the spectrum 
of  scalar cosmological perturbations 
generated from vacuum fluctuations during a slow-roll inflation phase.
Note that during slow-roll inflation, the Hubble parameter and the scalar 
field velocity slowly evolve: for a given scale, the above amplitude 
of the perturbations is determined by the value of $H$ and $\dot \phi$ when 
the scale exited the Hubble radius. Because of this effect, the obtained 
spectrum is not strictly scale-invariant. 

It is also instructive to recover the above result by a more 
 intuitive derivation.
One can think of the metric  perturbations in the radiation era as resulting 
from the time difference  for the end of inflation at different spatial 
points (separated by distances larger than the Hubble radius), 
the shift for the end of inflation being a consequence of the 
scalar field fluctuations  $\d\phi\sim H/2\pi$. 
Indeed,
\beq
\Psi_{\rm rad}\sim {\d a\over a}\sim H \d t, 
\eeq
and the time shift is related to the scalar field fluctuations 
by  $ \d t\sim \d\phi/\dot\phi$, which implies 
\beq
\Psi_{\rm rad}\sim {H^2\over \dot\phi},
\eeq
which agrees with the above result since during the radiation era 
$\R=(3/2)\Psi$ (see Eq.~(\ref{R_Psi2})).
It is also worth noticing that, during inflation, 
in the case of the slow-roll approximation, 
the term involving $C$ in the linear combination (\ref{v}) defining 
$v$ is negligible with respect to the term involving  $\d\phi$. 
One can therefore ``ignore'' the r\^ole of the metric perturbations
{\it during inflation} 
in the computation of the quantum fluctuations and consider only
the scalar field perturbations. But this simplification is valid only 
in the context of slow-roll approximation. It is not valid in the general 
case, as can be verified for inflation with a power-law scale factor.

\subsection{Gravitational waves}
We have focused so far  our attention on scalar perturbations, which 
are the most important in cosmology. Tensor perturbations, or primordial 
gravitational waves, if ever detected in the future, would be a remarkable 
probe of the early universe. In the inflationary scenario, 
like scalar perturbations, primordial 
gravitational waves are generated from vacuum quantum fluctuations.
 Let us now explain briefly this mechanism.

The action expanded at  second order in the perturbations contains a 
tensor part, which  
given by 
\beq
S^{(2)}_g={1\over 64\pi G}\int  d\eta\,  d^3x \, a^2 
\eta^{\mu\nu}\partial_\mu \bar E^i_j\partial_\nu\bar E^j_i,
\eeq
where $\eta^{\mu\nu}$ denotes the Minkoswki metric.
Apart from the tensorial nature of $E^i_j$, this action is quite 
similar to that of a scalar field in a FLRW universe (\ref{action_dS}), 
up to a renormalization 
factor $1/\sqrt{32\pi G}$.
The decomposition
\beq
a\bar E^i_j=\sum_{\lambda=+,\times}
\int {d^3k\over (2\pi)^{3/2}} 
 v_{k,\lambda}(\eta)\epsilon^i_j({\vec k};\lambda) e^{i \vec k.\vec x}
\eeq
where the $\epsilon^i_j({\vec k};\lambda)$ are the polarization tensors,
shows that the gravitational waves are essentially equivalent to two 
massless
scalar fields (for each polarization) $\phi_\lambda=m_P\bar E_\lambda /2$.

The total power spectrum is thus immediately deduced from (\ref{ps_phi}): 
\beq
{\cal P}_g=2\times {4\over \mP^2}\times \hbar\left({H\over 2\pi}\right)^2, 
\eeq
where the first factor comes from the two polarizations, the 
second from the renormalization with respect to a canonical scalar field, 
the last term being the power spectrum for a scalar field derived earlier.
In summary, the tensor power spectrum is
\beq
{\cal P}_g={2\hbar\over \pi^2}\left({H\over \mP}\right)_{k=aH}^2,
\label{power_T}
\eeq
where the label recalls that the Hubble parameter, which can be slowly evolving
during inflation, must be evaluated when the relevant scale exited the 
Hubble radius during inflation.

\subsection{Power spectra}
Let us rewrite the scalar and tensor power spectra, respectively 
given in (\ref{power_S}) and (\ref{power_T}), in terms of the 
scalar field potential only. This can be done by using the slow-roll 
equations 
(\ref{sr1}-\ref{sr2}). One finds for the scalar spectrum
\beq
{\cal P}_\R={1\over 12\pi^2}\left({V^3\over \mP^6{V'}^2}\right)_{k=aH}
\eeq
with subscript meaning  that the term 
on the right hand side must be evaluated at {\it Hubble radius exit}
for the scale of interest.
The scalar spectrum can also be written in terms of the 
first slow-roll parameter defined in (\ref{epsilon}), 
in which case it reads
\beq
{\cal P}_\R={1\over 24\pi^2}\left({V\over \mP^4\epsilon_V}\right)_{k=aH}.
\eeq
From the observations of the CMB fluctuations, 
\beq
{\cal P}_\R^{1/2}={1\over 2\sqrt{6}\pi}\left({V^{1/2}\over 
\mP^2\epsilon_V^{1/2}}\right)\simeq 5\times 10^{-5}.
\eeq
If $\epsilon_V$ is order $1$, as in chaotic models, one can evaluate 
the typical energy scale during inflation as
\beq
V^{1/4}\sim 10^{-3}\mP\sim 10^{15}{\rm GeV}.
\eeq

The tensor power spectrum, in terms of the scalar field potential, is 
given by
\beq
{\cal P}_g={2\over 3\pi^2}\left({V\over \mP^4}\right)_{k=aH}.
\eeq
The ratio of the tensor and  scalar amplitudes  is proportional to the 
slow-roll parameter $\epsilon_V$:
\beq
\label{r}
r\equiv
{{\cal P}_g\over {\cal P}_\R}=16\epsilon_V.
\eeq

The scalar and tensor spectra are almost scale invariant but not quite
since the scalar field evolves slowly during the inflationary phase.
In order to evaluate quantitatively this variation, it is convenient to
introduce a scalar  {\it spectral index} as well as a tensor one, 
defined respectively by  
\beq
n_S(k)-1={d\ln {\cal P}_\R(k)\over d\ln k}, \quad 
n_T(k)={d\ln {\cal P}_g(k)\over d\ln k}.
\eeq
One can express the spectral indices in terms of the slow-roll 
parameters. For this purpose, let us note that, in the slow-roll 
approximation,  $d\ln k=d\ln(aH)\simeq d\ln a$, so that  
\beq
{d\phi\over d\ln a}=
{\dot\phi\over H}\simeq -{V'\over 3H^2}\simeq - m_P^2{V'\over V},
\eeq
where the slow-roll equations 
 (\ref{sr1}-\ref{sr2}) 
have been used. 
Therefore, one gets 
\beq
n_s(k)-1=2\eta_V- 6 \epsilon_V,
\eeq
where $\epsilon_V$ and $\eta_V$ are the two slow-roll parameters defined
in  (\ref{epsilon}) and (\ref{eta}). 
Similarly, one finds for the tensor spectral index
\beq
n_T(k)=-2\epsilon_V.
\eeq
Comparing with Eq.~(\ref{r}), this yields the 
relation
\beq
r=-8 n_T,
\eeq
the  so-called {\it consistency
relation}
which relates  purely {\it observable}  quantities. This means 
that if one was able to observe the primordial gravitational 
waves and  measure the amplitude and spectral index of their
spectrum, a rather  formidable task, then one 
would be able to test directly the paradigm of single field slow-roll
inflation.

Finally, let us mention the possibility to get information 
on inflation from the measurement of the running 
of the spectral index.
Introducing  the second-order slow-roll parameter
\beq 
\xi_V = \mP^4 \frac{V'V'''}{V^2} , 
\eeq 
the running is given by 
 \beq \frac{d
n_s}{d \ln k} = -24 \epsilon_V^2 +16 \epsilon_V \eta_V -2 \xi_V .
\eeq
As one can see, the amplitude of the variation depends on the slow-roll
parameters and thus on the models of inflation. 

\subsection{Conclusions}
To conclude, let us mention the existence of 
more sophisticated models of inflation, such as models where 
several scalar fields contribute to inflation. 
In contrast with the single inflaton case, 
which  can generate only adiabatic primordial fluctuations, because
all types of matter are decay product of the same inflaton, 
{\it multi-inflaton models} can generate both adiabatic and isocurvature
perturbations, which can even be correlated 
\cite{l99}. 

Another recent direction of research is   the possibility to 
disconnect the  fluctuations of the inflaton, the field that drives
inflation, from the observed cosmological perturbations, which 
could have  been generated 
from the quantum fluctuations of another scalar field 
\cite{curvaton}.

From  the theoretical point of view an
 important challenge  remains 
 to  identify  viable and natural candidates for the 
inflaton field(s) in the framework of high energy physics, with the hope
that future observations of the  cosmological perturbations 
will be precise enough to discriminate between  various candidates 
and thus give us a clue about  which physics really drove inflation.

Alternative scenarios to inflation can also been envisaged, as long as 
they can predict unambiguously  primordial fluctuations compatible
with the present observations. 
In this respect,  one must emphasize  that the cosmological perturbations 
represent today essentially  the only observational window that gives access 
to the very high energy physics, hence the importance for any 
early universe model to be able to give firm predictions 
for  the primordial fluctuations it generates.

\vskip 1cm
{\bf Acknowledgements:}

I would like to thank the organizers of this Cargese school for inviting me 
to lecture at this very nice village in Corsica. I also wish to thank 
the other lecturers and participants for their remarks and questions, as 
well as G\'erard Smadja and 
Filippo Vernizzi for their careful reading of the draft.

\end{document}